\documentclass[10pt,leqno]{amsart}
\usepackage{graphicx}
\usepackage{indentfirst,csquotes}

\usepackage{amssymb,amsthm,amsmath}
\usepackage{xcolor,paralist,hyperref,fancyhdr,etoolbox}

\hypersetup{ colorlinks=true, linkcolor=black, filecolor=black, urlcolor=black }

\usepackage{cite}
\usepackage{algorithmic}
\usepackage{flushend}
\usepackage{textcomp}
\usepackage{subcaption}

\begin{document}
\title{Resource-Aware Model Selection for \\Scalable Indoor Localization on HPC Platforms} 

\author[Fukuharu Tanaka]{Fukuharu Tanaka}
\date{\today}
\address{RIKEN Center for Computational Science, Kobe, Hyogo, Japan}
\email{f-tanaka@ist.osaka-u.ac.jp}
\author[Hamada Rizk]{Hamada Rizk}
\date{\today}
\address{The University of Osaka, Suita, Osaka, Japan}
\email{hamada\_rizk@ist.osaka-u.ac.jp}
\author[Moustafa Youssef]{Moustafa Youssef}
\date{\today}
\address{The American University in Cairo, New Cairo, Egypt}
\email{moustafa-youssef@aucegypt.edu}
\author[Hirozumi Yamaguchi]{Hirozumi Yamaguchi}
\date{\today}
\address{The University of Osaka, Suita, Osaka, Japan}
\email{h-yamagu@ist.osaka-u.ac.jp}

\begin{abstract}
Large-scale indoor localization is increasingly needed in campuses, smart buildings, factories, and digital-twin infrastructures, where wireless conditions, access-point deployments, and spatial layouts evolve over time. Such systems must be accurate, extendable, and maintainable, allowing new buildings, floors, rooms, and service areas to be added without retraining a monolithic model. Modular learning-based localization supports this goal by assigning independent models to buildings, floors, and fine-grained spatial regions. However, this extendability introduces a high-performance inference challenge: each query may require selecting and executing among hundreds or thousands of local models, making exhaustive inference costly in computation, accelerator memory residency, model loading, and scheduling.
This paper presents a resource-aware modular inference framework for WiFi fingerprint-based indoor localization on high-performance and distributed computing platforms. The framework organizes local autoencoder models into a building-floor-spot hierarchy and formulates localization as model selection over a large pretrained model ensemble. To reduce inference cost under resource constraints, we introduce two lightweight execution-pruning strategies: Hierarchical Candidate Pruning, which performs coarse-to-fine model selection, and Trajectory-Aware Pruning, which uses temporal locality in user movement to restrict inference to spatially plausible neighboring models.
Experiments on a real-world dataset demonstrate that the proposed framework delivers scalable inference without sacrificing localization quality. Compared with exhaustive evaluation over 735 spot models, Hierarchical Candidate Pruning requires only 67 model evaluations, while Trajectory-Aware Pruning reduces this number to just 10, cutting model executions by 98.6\%. 
Under constrained-memory execution, the same pruning strategies also lower latency by minimizing inactive model loading, highlighting their potential for extendable, high-performance, and distributed indoor localization services.
Furthermore, evaluations on the Fugaku supercomputer demonstrate that our resource-aware pruning achieves high parallel scalability and substantially lowers inference latency in multi-node HPC environments.
\end{abstract} 

\maketitle

\footnotetext{© 2026 IEEE. Personal use of this material is permitted.
Permission from IEEE must be obtained for all other uses, in any current
or future media, including reprinting/republishing this material for
advertising or promotional purposes, creating new collective works,
for resale or redistribution to servers or lists, or reuse of any
copyrighted component of this work in other works.}




\section{Introduction}

Indoor localization is a key enabling technology for smart buildings, campuses, factories, hospitals, and digital-twin services, where GPS is unavailable and location awareness is required for navigation, emergency response, asset tracking, and context-aware automation\cite{11167372, mostafa2025survey, aziz2025comprehensive}. As these environments expand in size and complexity, localization systems must satisfy requirements beyond positioning accuracy. They must remain extendable when new areas are added, maintainable when local wireless conditions change, and efficient enough to support repeated inference under practical computing constraints.
In particular, for centralized positioning services handling concurrent requests across large facilities, backend HPC platforms require high inference throughput and scalable model management.

WiFi fingerprinting remains one of the most practical approaches for indoor localization because it reuses already deployed wireless infrastructure and can be supported by commodity mobile devices\cite{zholamanov2025rssi, dai2023survey, neupane2025wi}. 
By learning the relationship between received signal strength patterns and indoor locations, fingerprinting methods can capture complex propagation effects such as attenuation, reflection, and multipath, which are difficult to model analytically. These advantages have made WiFi fingerprinting a widely used solution for multi-building and multi-floor indoor positioning scenarios.

However, conventional learning-based fingerprinting systems are often designed around a monolithic model that represents the entire deployment area\cite{feng2022survey,wang2020indoor,wang2015deepfi,wang2015phasefi,chen2017confi,song2019cnnloc,kim2018scalable,ayinla2026reloc}. 
Such a design is simple to train and deploy in small, fixed environments, but it becomes increasingly problematic at scale. When a new building, floor, room, or service area is introduced, or when access points are replaced and local propagation patterns change, the global model may require expensive retraining. Moreover, as the number of reference locations grows, the output space becomes larger, the learning problem becomes more difficult, and inference over the full localization space becomes more costly. This limits the practicality of monolithic localization models for long-term operation in evolving indoor environments.

A modular alternative is to decompose the environment into smaller spatial units and assign independent models to them\cite{abbas2019wideep,  saeed2022cellstory}. 
This design fundamentally improves operational maintainability for long-term deployment: when wireless conditions or room layouts change in a localized area, only the affected local model needs to be updated. 
Unrelated spot models and higher-level spatial models remain untouched unless facility-wide propagation patterns shift. Consequently, routine environmental maintenance avoids expensive, full-system retraining. 
Nevertheless, modularity shifts the main challenge from model training to model execution. 
A large deployment may contain hundreds or thousands of local models, and exhaustive evaluation of all models for every query is inefficient. In accelerator-based systems, the bottleneck is not only neural-network computation, but also model residency in GPU memory, model loading, cache replacement, and scheduling. In distributed or multi-node settings, model placement and repeated loading can further affect throughput and resource contention. Therefore, scalable modular localization requires resource-aware inference mechanisms that select a small but reliable active model set for each query.

This paper addresses this systems challenge by formulating modular indoor localization as a resource-aware model-selection problem. Instead of executing all local models, the proposed framework organizes the environment into a building-floor-spot hierarchy and dynamically reduces the candidate model set before inference. Each spatial unit is represented by an autoencoder trained to reconstruct WiFi fingerprints from its corresponding region. During inference, reconstruction error is used as a similarity score, while candidate pruning determines which models should be evaluated under compute and memory constraints.

We introduce two complementary pruning strategies. Hierarchical Candidate Pruning (HCP) performs coarse-to-fine model selection by first identifying likely buildings, then floors, and finally spot-level candidates. Trajectory-Aware Pruning (TAP) exploits temporal locality in human movement by restricting the active model set to spatially plausible neighbors of the previously estimated spot. These strategies are designed not only to reduce arithmetic computation, but also to lower model-loading overhead and improve suitability for high-performance and distributed inference environments.

The proposed framework is evaluated using the UJIIndoorLoc dataset, a real-world WiFi fingerprinting benchmark covering multiple buildings and floors. The results show that resource-aware pruning can significantly reduce inference cost while preserving localization quality. Compared with exhaustive evaluation over 735 spot models, HCP reduces the number of evaluated models to 67 with comparable top-1 accuracy, while TAP reduces the active set to only 10 models, achieving a 98.6\% reduction in model executions. Under constrained-memory execution, the proposed pruning strategies also reduce latency by avoiding repeated loading of inactive models. These results indicate that modular indoor localization can be made more practical for large-scale deployment when model selection, memory residency, and execution cost are considered as first-class system concerns.
Furthermore, scaling experiments on the Fugaku supercomputer confirm that suppressing model evaluations directly compresses the dominant neural-network execution time, achieving linear latency reduction in multi-node CPU environments.

The main contributions of this study are fourfold: 
(1) we formulate modular indoor localization as a resource-aware inference problem over a large ensemble of local models, where execution cost, memory residency, and model loading are key constraints; 
(2) we propose a hierarchical building--floor--spot localization framework with local autoencoder models, enabling incremental extension and localized updates; 
(3) we introduce two execution-pruning strategies, Hierarchical Candidate Pruning (HCP) and Trajectory-Aware Pruning (TAP), to reduce the active model set using spatial hierarchy and temporal locality; and 
(4) we evaluate the framework on the UJIIndoorLoc dataset and validate its parallel scalability on the Fugaku supercomputer, demonstrating that HCP and TAP reduce model executions from 735 to 67 and 10, respectively, and achieve ultra-low inference latency down to 1.07 ms across 16 CPU nodes while maintaining competitive accuracy (e.g., 4.04 m mean distance error for HCP).

\section{Related Works}

\subsection{Sensor-Based Indoor Localization}

With the rapid spread of smart buildings, underground complexes, and large commercial facilities, the importance of location-based services in indoor environments where GPS signals are unavailable has grown substantially\cite{11167372, mostafa2025survey, aziz2025comprehensive}.
To address this need, sensor-based localization techniques have long been studied as a means of achieving high-precision indoor positioning by deploying dedicated sensing hardware inside buildings\cite{dicu2025comprehensive, roy2022survey}.

Representative examples include time-of-arrival localization using ultra-wideband (UWB) \cite{saleh2026vehicular, trinh2025uwb}, angle-of-arrival and received-signal-strength localization using visible light communication (VLC) \cite{cai2023indoor, zhu2024survey}, and proximity-based localization using Bluetooth Low Energy (BLE) beacons\cite{kotrotsios2022design, shi2024survey}.
Vision-based systems using cameras and pedestrian dead reckoning based on smartphone inertial measurement units (IMUs) have also been investigated \cite{kwon2023feasibility}.

However, these sensor-based approaches fundamentally suffer from very high installation and maintenance costs.
Systems based on UWB, VLC, or BLE require dense deployment of hardware such as anchors, LEDs, or beacons on ceilings and walls at intervals of only a few to several tens of meters.
This not only incurs substantial equipment costs but also leads to significant installation overhead, including wiring, mounting work, and system integration.
In addition, battery-powered devices such as BLE beacons require periodic replacement, while high-precision radio systems such as UWB often require recalibration when the indoor environment changes due to furniture movement or structural modification.
Vision-based systems also face practical limitations, including lighting conditions, line-of-sight constraints, and privacy concerns, which make large-scale deployment difficult.

\textit{These limitations motivate localization frameworks that can reuse existing wireless infrastructure while remaining practical for large and evolving indoor environments. The proposed framework addresses this gap by relying on WiFi fingerprints rather than dedicated sensing hardware, while focusing on scalability and maintainability.}

\subsection{Fingerprint-Based Localization}

To overcome the economic burden of dedicated infrastructure, fingerprint-based localization methods using received signal strength indicator (RSSI) or channel state information (CSI) from existing WiFi access points and cellular base stations have been widely adopted \cite{zholamanov2025rssi, dai2023survey, neupane2025wi}.
Early fingerprinting methods were dominated by deterministic or probabilistic pattern-matching approaches such as RADAR\cite{bahl2000radar} and Horus\cite{youssef2005horus, youssef2003wlan, el2011impact}.
More recently, deep learning-based models have been proposed to better capture complex multipath fading and attenuation caused by walls and dynamic obstacles in indoor environments\cite{feng2022survey, wang2020indoor}.
Representative examples include DeepFi, which learns CSI fingerprints with a deep belief network\cite{wang2015deepfi}, PhaseFi, which uses calibrated CSI phase information\cite{wang2015phasefi}, and ConFi, which applies convolutional neural networks to CSI-based WiFi localization\cite{chen2017confi}.
On the RSSI side, CNNLoc combines stacked autoencoders with a one-dimensional CNN for multi-building and multi-floor localization\cite{song2019cnnloc}, while the stacked autoencoders (SAE) exploit the hierarchical structure of building and floor estimation to support multi-building and multi-floor indoor localization\cite{kim2018scalable}.

Recent studies have further improved fingerprint-based localization through multimodal sensing, efficient learning, and robustness to deployment changes.
RRLoc and SelfLoc combine RSSI and RTT to exploit complementary signal information, with SelfLoc additionally using self-supervised learning to reduce the need for labeled data\cite{rizk2022robust, rizk2025selfloc}.
MambaLoc further explores efficient CSI localization using Mamba with cross-modal knowledge distillation from UWB measurements\cite{bahnassy2025mambaloc}.
Other studies have addressed practical deployment challenges: GlobLoc improves cross-environment generalization without recalibration, while LiPhi++ reduces the effort required to construct and maintain fingerprint databases\cite{rizk2023indoor, rizk2023laser}.
For multi-floor localization, CellRise exploits sequential cellular signals to learn floor-discriminative representations\cite{rizk2020ubiquitous}, while GraphLy models spatial relationships with a graph neural network to improve robustness in cluttered environments\cite{rizk2024adaptability}.

A major limitation of these deep fingerprinting methods is that they typically represent the entire spatial area with a single global model.
In such monolithic designs, all buildings, floors, or reference locations are jointly learned by one neural network, either as a multi-class classification problem or as a coordinate regression problem.
Although this is convenient in controlled settings, it becomes highly fragile in long-term real deployments.
Indoor environments are dynamic, and office layout changes, access point failures or replacements, and propagation variations occur frequently.
In campus- or commercial-scale environments, new floors or new buildings may also be added over time.
Under such conditions, even a small local change can degrade the performance of the global model because the learned parameters are tightly coupled with the entire spatial space.
As a result, practical operation often requires re-collecting fingerprints over the whole environment and retraining the entire model from scratch, even when only a small region has changed.

\textit{This gap motivates a shift from monolithic fingerprinting models to modular localization architectures that support localized updates and incremental extension. The proposed framework follows this direction by decomposing the localization space into building-, floor-, and spot-level models rather than relying on a single global model.}

\subsection{Modular Model Decomposition in Fingerprint-Based Localization}

To address the maintenance and extensibility limitations of monolithic fingerprinting models, modular localization approaches have been proposed in which the environment is decomposed into smaller spatial units and an independent model is assigned to each unit.
A representative example is WiDeep~\cite{abbas2019wideep}, which constructs a separate deep model for each fingerprint reference point and combines their outputs in a probabilistic framework.
Another notable example is CellStory~\cite{saeed2022cellstory}, which trains a distinct stacked denoising autoencoder for each floor and estimates the target floor based on reconstruction error.
These modular designs improve maintainability because local environmental changes can be handled by updating only the affected model, without retraining the entire system.

However, modular decomposition also introduces a new scalability challenge at inference time.
As the environment grows to building-scale, campus-scale, or city-scale deployments, the number of local models increases rapidly.
In such settings, evaluating all candidate models for every query becomes computationally expensive and may also impose substantial memory and scheduling overhead.
As the number of modeled spots increases, brute-force evaluation of all local models becomes increasingly impractical for real-time localization.
Therefore, although modular approaches are attractive from the perspective of maintenance and extensibility, they require an additional mechanism for reducing the number of candidate models at inference time.

\textit{The proposed framework fills this gap by treating modular localization as a resource-aware inference problem. It introduces hierarchical and trajectory-aware candidate pruning to reduce model executions, model-loading overhead, and accelerator-memory pressure while preserving the benefits of modular model decomposition.}

\section{System Overview}

\begin{figure}[t]
    \centering
    \includegraphics[width=0.7\linewidth]{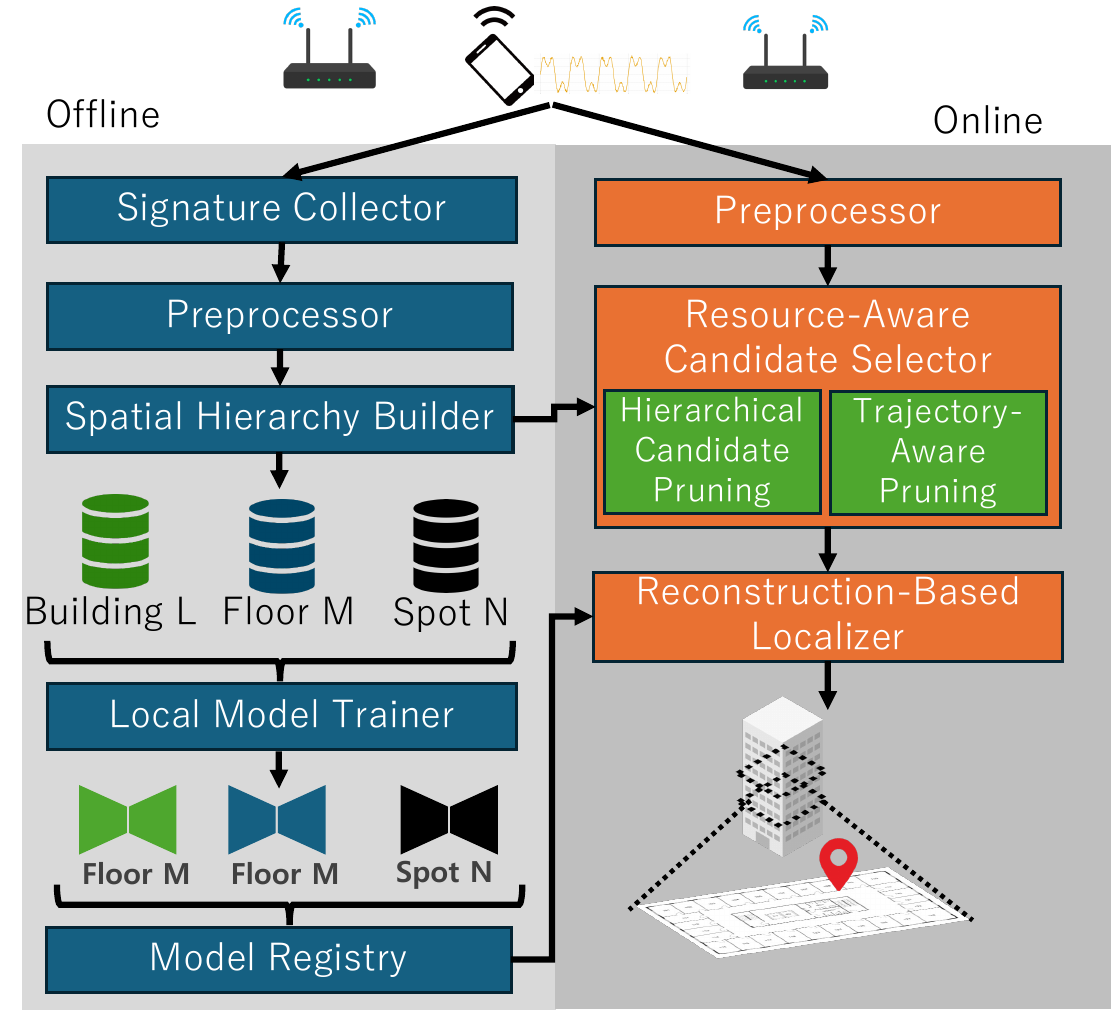}
    \caption{Proposed Modular Indoor Localization Architecture.}
    \label{fig:SysA}
\end{figure}

Figure~\ref{fig:SysA} shows the proposed system architecture. The system runs in two phases: the offline training phase and the online localization phase. During the offline phase, the system constructs a set of local deep models corresponding to the spatial hierarchy of the target environment. 
Instead of training one monolithic model for the entire area, the environment is decomposed into buildings, floors, and spots, and independent autoencoder models are trained for these spatial units. 
Conceptually, these local autoencoders act as novelty or anomaly detectors: each model is trained exclusively on fingerprints collected within its designated spatial unit, learning a tight representation of that region's signal distribution. 
Consequently, when presented with a query fingerprint from its own region, the model yields a low reconstruction error, whereas out-of-distribution fingerprints from other regions produce significantly higher errors. This generalization gap provides the foundational mechanism for using reconstruction error as a similarity score for model selection.
This modular design improves extendability, since new areas can be added by training new local models, and improves maintainability, since only the models affected by local environmental changes need to be updated.

To collect the training data, the \textbf{Signature Collector} module scans the available WiFi access points and records their received signal strengths at different locations in the area of interest. These fingerprints are then forwarded to the \textbf{Preprocessor} module, which formats the RSS vectors, handles unavailable access points, and applies the required normalization before training. The preprocessed fingerprints are passed to the \textbf{Spatial Hierarchy Builder} module, which assigns each sample to its corresponding building, floor, and spot. Based on this hierarchy, the \textbf{Local Model Trainer} module trains independent autoencoder models for the different spatial units. Building-level models capture coarse wireless patterns, floor-level models capture vertical and floor-specific characteristics, and spot-level models learn fine-grained local fingerprint distributions. Finally, the trained models and their spatial metadata are stored in the Model Registry, which is later used to retrieve and execute only the models required during online localization.
During the online localization phase, the user is localized in real time. The process starts by scanning the WiFi access points and their RSS values at the unknown user location. The collected fingerprint is first processed by the \textbf{Preprocessor} module to match the input format used during training. 
Because obtaining the reconstruction error for any candidate model requires executing a full model forward pass for the specific query, executing all local models per query creates a major bottleneck in execution time, memory residency, and model loading. 
To overcome this, the \textbf{Resource-Aware Candidate Selector} module dynamically determines a small, promising subset of candidate models to be evaluated for the current query, thereby pruning unnecessary model forward executions. 
It uses the \textbf{Hierarchical Candidate Pruning} sub-module to narrow the search from building to floor and then to spot. When a recent previous estimate is available, the \textbf{Trajectory-Aware Pruning} sub-module further restricts the candidate set to spatially plausible neighboring spots. This reduces unnecessary model executions and limits model loading under accelerator-memory or distributed execution constraints.
The selected candidate models are then passed to the \textbf{Reconstruction-Based Localizer} module. Each candidate autoencoder reconstructs the input fingerprint, and the reconstruction error is used as a similarity score between the input scan and the spatial unit represented by the model. The spot corresponding to the model with the lowest reconstruction error is returned as the estimated user location. By separating offline model construction from online resource-aware model selection, the proposed system supports scalable, extendable, and efficient indoor localization for high-performance and distributed inference environments.

\section{Proposed Method}

\subsection{Problem Formulation}

Let a WiFi fingerprint be represented by an $M$-dimensional RSSI vector
\begin{equation}
x \in \mathbb{R}^{M}
\end{equation}
where $M$ denotes the number of access points used to construct the fingerprint representation after preprocessing.
Let the set of candidate spots be
\begin{equation}
\mathcal{S} = \{s_1, s_2, \dots, s_N\}
\end{equation}
where each spot corresponds to a spatial unit such as a room, a corridor segment, or another localized region in the environment.

For each candidate spot $u \in \mathcal{S}$, the corresponding local autoencoder takes the query fingerprint $x$ as input and produces a reconstructed fingerprint $\hat{x}^u$ through a full forward pass of its encoder ($\text{Enc}_u$) and decoder ($\text{Dec}_u$):
\begin{equation}
\hat{x}^u = \text{AE}_u(x) = \text{Dec}_u(\text{Enc}_u(x)).
\end{equation}
The reconstruction error is then computed as
\begin{equation}
e_u(x) = \|x - \hat{x}^u\|_2^2.
\end{equation}
The localization problem is formulated as selecting the candidate spot that minimizes this reconstruction error:
\begin{equation}
\hat{u} = \arg\min_{u \in \mathcal{S}} e_u(x).
\end{equation}

This reconstruction-error-based formulation enables fine-grained localization while preserving scalability and maintainability through modular model decomposition.

\subsection{Offline-Phase}

\subsubsection{Preprocessor}
This module converts the recorded WiFi RSSI readings $x_i$ into feature vectors used by the learning model. 
Since not all of the $M$ access points installed in the target area are observable in every scan, the module assigns a default weak RSSI value of -100 dBm to any AP that is not detected.
The resulting RSSI values are then normalized to the range [0,1] on a per-AP basis.

\subsubsection{Spatial Hierarchy Builder}

This module constructs the spatial index that links the fingerprint dataset to the modular inference engine. 
Given the training fingerprints and their location labels, this module builds a metadata table in which each spot is represented by a unique identifier, its building and floor membership, and a representative coordinate computed from the samples assigned to that spot. 
In our implementation, the representative coordinate is defined as the centroid of the spot samples, which provides a compact geometric proxy for estimating inter-spot proximity.
This metadata table serves three purposes. 
First, it partitions the training data into building-, floor-, and spot-level subsets, enabling independent training of local autoencoder models at different spatial resolutions. 
Second, it encodes the parent--child relations among buildings, floors, and spots, allowing the inference engine to traverse the localization space in a coarse-to-fine manner. 
Third, it supports neighborhood queries by computing spatial proximity between spot centroids, which is required for trajectory-aware candidate selection.
During online inference, this spatial index is used as a lightweight control structure before neural model execution. 
Hierarchical Candidate Pruning uses the building--floor--spot relations to restrict the active model set to candidates that are spatially consistent with coarse-level predictions. 
Trajectory-Aware Pruning uses the centroid-based neighborhood structure to retrieve nearby spots around the previously estimated location. 
Thus, the Spatial Hierarchy Builder transforms raw location labels into an execution-aware spatial representation that enables scalable model selection, reduces unnecessary model evaluations, and supports localized updates when the environment changes.

\subsubsection{Local Model Trainer}

\begin{table}[t]
\centering
\caption{Training hyperparameters and network architecture.}
\label{tab:hyperparameters}
\begin{tabular}{ll}
\hline
Parameter & Value \\
\hline
Batch size & 256 \\
Maximum epochs & 100 \\
Learning rate & $1\times10^{-3}$ \\
Optimizer & Adam \\
Dropout rate & 0.2 \\
Early stopping patience & 8 epochs \\
Random seed & 42 \\
Encoder & 200--300--400--500 \\
Decoder & 500--400--300--200 \\
\hline
\end{tabular}
\end{table}

\begin{figure}[t]
    \centering
    \includegraphics[width=0.7\linewidth]{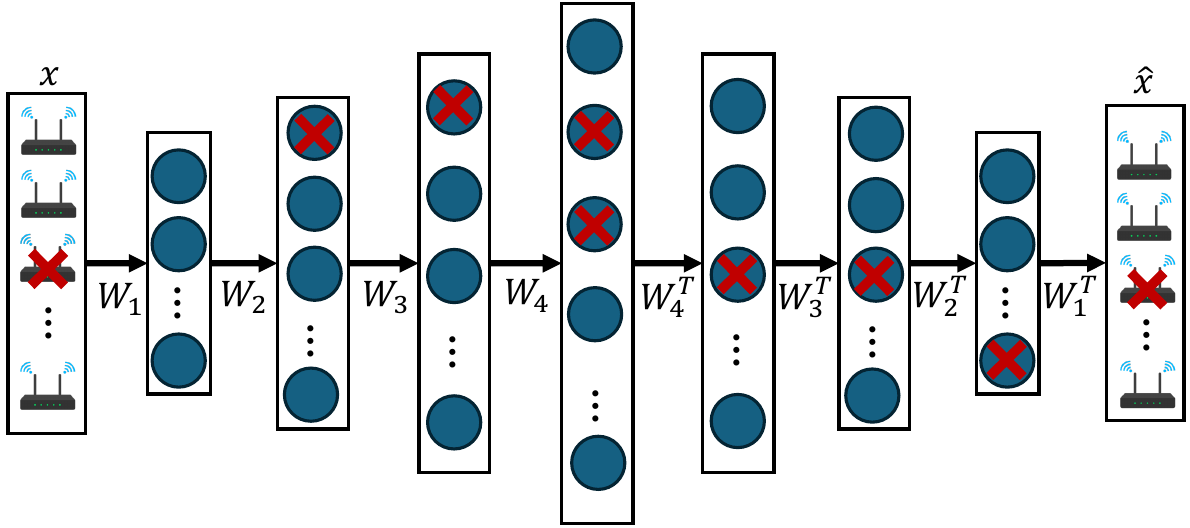}
    \caption{Model Architecture. Crossed nodes are example of dropped-out
neurons. Crossed WAPs are not detected.}
    \label{fig:model}
\end{figure}

The proposed framework consists of three levels of local models, namely building models, floor models, and spot models. 
Each model is trained independently using fingerprints belonging only to its corresponding spatial unit.

For a given target unit $u$, we train an autoencoder (AE) that reconstructs the input fingerprint as shown in Figure~\ref{fig:model}. 
Let $\hat{\mathbf{x}}_u$ denote the reconstructed output produced by the model for unit $u$. 
For both training and inference, the reconstruction loss is computed only over WAPs that are actually detected in the input fingerprint. 
Let $m_i \in \{0,1\}$ denote a binary mask for the $i$-th WAP, where $m_i=1$ if the corresponding RSSI value is observed and $m_i=0$ otherwise. 
Then, the reconstruction loss for unit $u$ is defined as
\begin{equation}
\mathcal{L}_u
=
\frac{\sum_{i=1}^{M} m_i \left( x_i - \hat{x}_{u,i} \right)^2}
{\sum_{i=1}^{M} m_i + \epsilon}.
\end{equation}
This masked formulation prevents the reconstruction score from being dominated by the large number of absent AP entries and focuses the model on the informative WAP dimensions that are actually observed.
The model is optimized so that fingerprints collected inside the target unit are reconstructed with smaller error than fingerprints from other units. 
This reconstruction-based design is suitable for spot-level localization because it does not require direct classification over a large global label space. 
Instead, each model learns the signal characteristics of one local region.

A key advantage of this modular formulation is that each spot model can be trained independently using only the data of its own spot, without using fingerprints from other spots. 
As a result, when a spot is added or updated, only the corresponding local model needs to be trained or retrained, which makes the framework more maintainable and extensible.

The AE architecture used in this work is summarized in Table \ref{tab:hyperparameters}. 
We use a fully connected encoder--decoder network with hidden dimensions 200--300--400--500 in the encoder and 500--400--300--200 in the decoder. 
The model is trained with a batch size of 256 for up to 100 epochs using Adam with a learning rate of $1 \times 10^{-3}$. 
Dropout with rate 0.2 is applied for regularization, and early stopping with patience 8 is used to avoid overfitting.

\subsection{Online-Phase}

During online inference, the input fingerprint is first preprocessed, after which the candidate selection module determines whether HCP or TAP is applied. 
The selected candidate set is then passed to the reconstruction-based localizer.

\subsubsection{Resource-Aware Candidate Selector}
This module selects the candidate-reduction strategy to be applied at inference time according to the availability and reliability of past location estimates. 
When no valid previous estimate is available, or when the temporal gap is too large to assume local movement continuity, the system applies HCP to obtain a coarse-to-fine candidate set. 
When a recent previous spot estimate is available, the system applies TAP to restrict the search to spatially nearby spots and reduce the number of models to be evaluated. 
For recursive inference, if repeated TAP updates are expected to accumulate localization errors, the selector can fall back to HCP to reinitialize the search from a broader spatial context.

\subsubsection{Hierarchical Candidate Pruning}

\begin{figure}[t]
    \centering
    \includegraphics[width=0.7\linewidth]{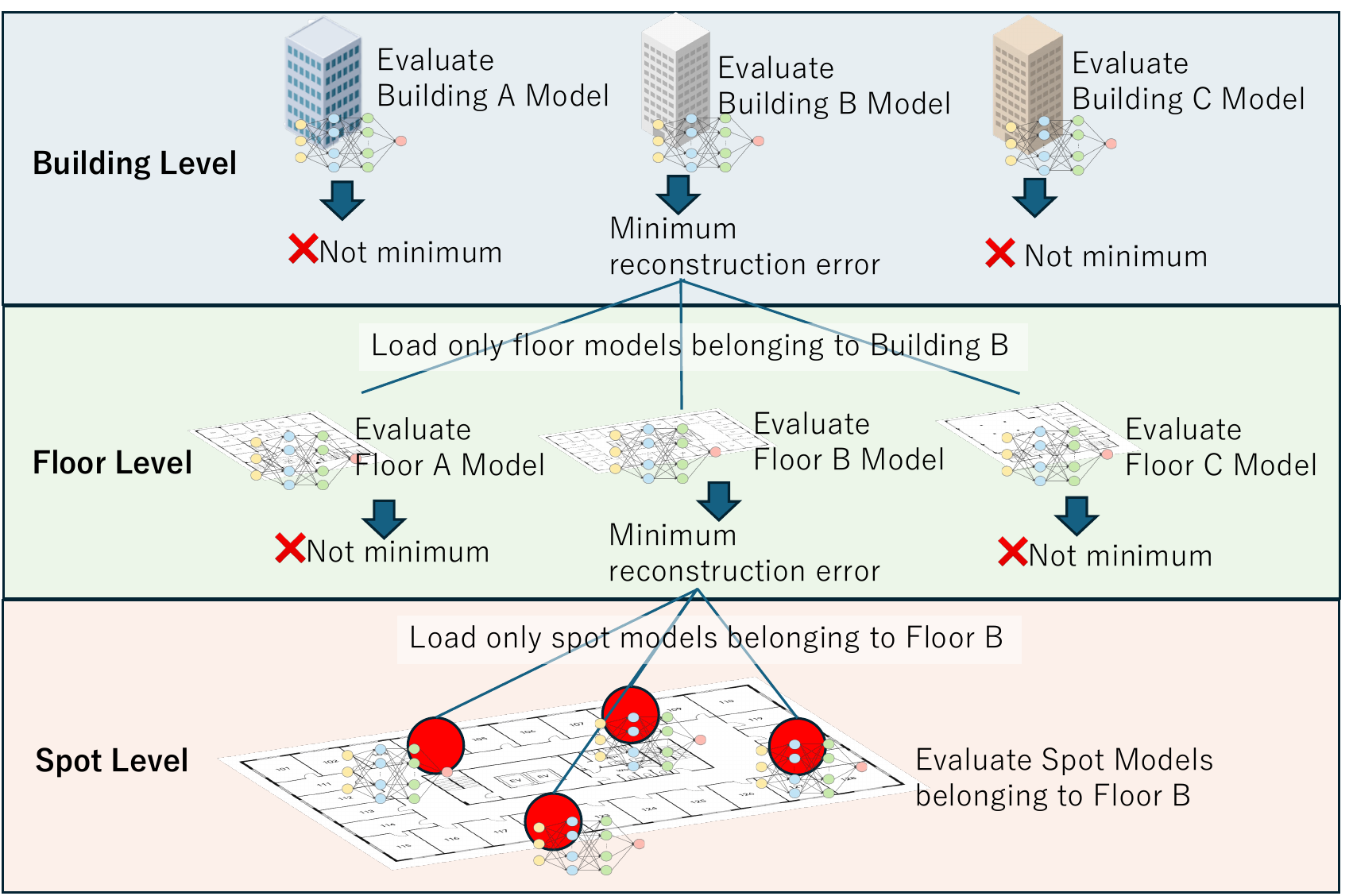}
    \caption{Hierarchical Candidate Pruning Workflow.}
    \label{fig:HCP}
\end{figure}

Evaluating all spot models at every inference step is computationally expensive, especially in large indoor environments. 
To avoid exhaustive search, the proposed framework exploits the spatial hierarchy of the environment. The inference procedure is performed in three stages as shown in Figure~\ref{fig:HCP}.

First, the system evaluates all building models and selects the $K_b$ buildings with the smallest reconstruction errors. Let $\mathcal{B}$ denote the set of all building candidates, and let $\mathcal{B}^*$ denote the selected subset:
\begin{equation}
\mathcal{B}^* = \operatorname{TopK}_{b \in \mathcal{B}} \bigl(-e_b(\mathbf{x})\bigr),
\end{equation}
where $e_b(\mathbf{x})$ is the reconstruction error of the building model for building $b$.

Second, only the floor models belonging to the selected buildings are evaluated. Let $\mathcal{F}(b)$ denote the set of floors in building $b$. The selected floor candidates are given by
\begin{equation}
\mathcal{F}^* =
\bigcup_{b \in \mathcal{B}^*}
\operatorname{TopK}_{f \in \mathcal{F}(b)} \bigl(-e_{b,f}(\mathbf{x})\bigr),
\end{equation}
where $e_{b,f}(\mathbf{x})$ is the reconstruction error of the floor model for floor $f$ in building $b$.

Third, only the spot models belonging to the selected floors are considered. 
Let $\mathcal{S}(b,f)$ denote the set of spots in floor $f$ of building $b$. 
The final spot candidates are obtained as
\begin{equation}
\mathcal{S}^* =
\bigcup_{(b,f)\in \mathcal{F}^*}
\operatorname{TopK}_{s \in \mathcal{S}(b,f)} \bigl(-e_{b,f,s}(\mathbf{x})\bigr),
\end{equation}
where $e_{b,f,s}(\mathbf{x})$ is the reconstruction error of the spot model associated with spot $s$.

It should be noted that including coarse-level models in HCP introduces a hierarchical maintenance trade-off. 
While spot-level models provide strict update locality (requiring retraining only for the modified spot), floor and building models aggregate signal patterns across larger spatial scopes. Consequently, while minor local spot changes do not impact higher-level models, broader wireless reconfigurations may require updating the corresponding floor or building model. Nevertheless, because such retraining remains bounded strictly to the affected branch of the spatial hierarchy, environmental updates still avoid full-system retraining, preserving the localized maintenance advantage over monolithic approaches.

\subsubsection{Trajectory-Aware Pruning}

\begin{figure}[t]
    \centering
    \includegraphics[width=0.7\linewidth]{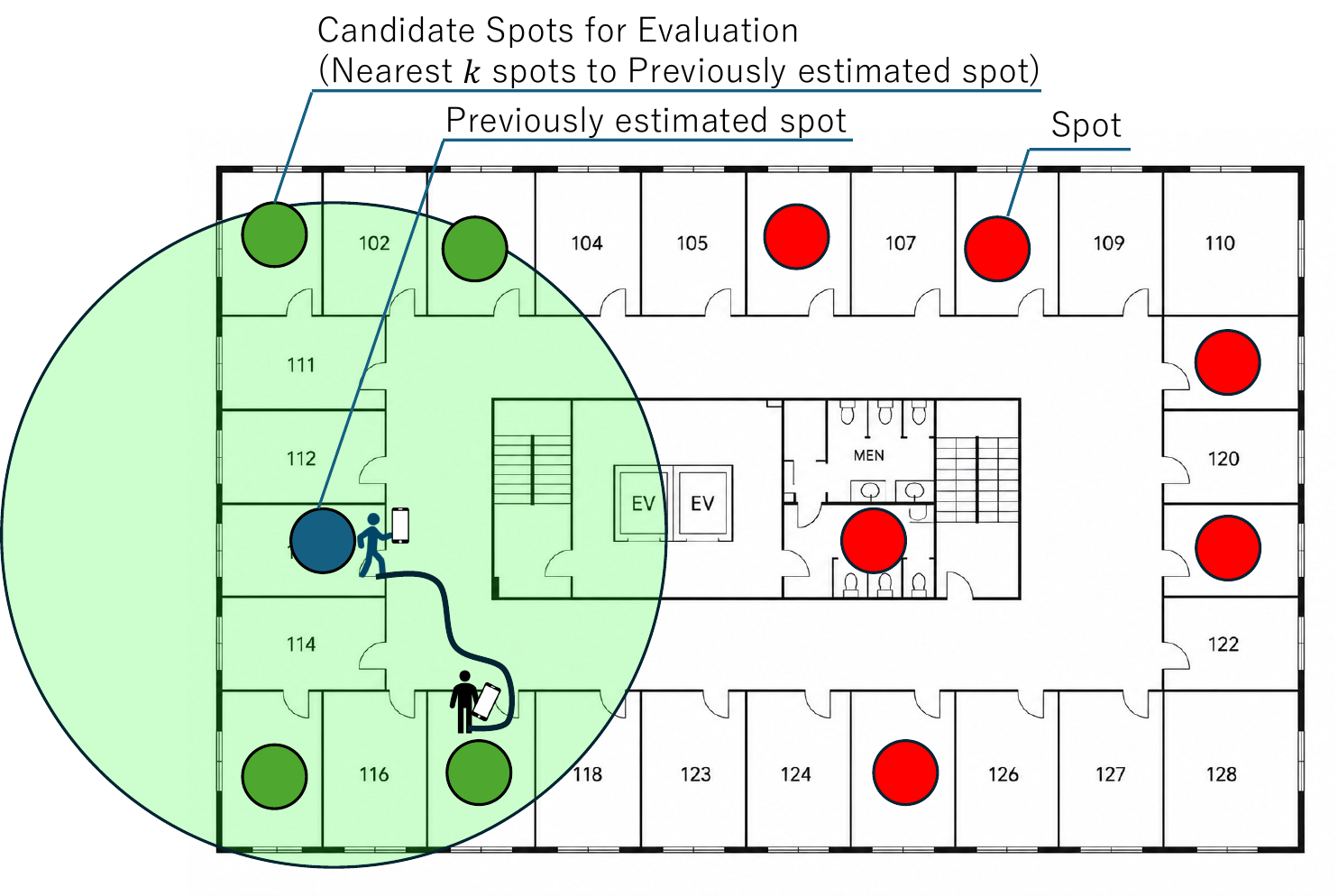}
    \caption{Trajectory-Aware Pruning Workflow.}
    \label{fig:TAP}
\end{figure}

The proposed method further reduces inference cost by exploiting the continuity of human movement. 
In normal indoor mobility, a user does not instantly move from one distant building to another without passing through intermediate spots. 
Therefore, when a previously estimated location is available and the time gap is sufficiently small, the previous spot ID can be used as a reference point to narrow the candidate set for the next inference step, as shown in Figure~\ref{fig:TAP}.

Let $\hat{s}_{t-1}$ be the previously estimated spot ID at time $t-1$. We define the neighborhood of $\hat{s}_{t-1}$ as the nearest $K$ spots to the previous estimate:
\begin{equation}
\mathcal{N}(\hat{s}_{t-1}) = \operatorname{TopK}_{s \in \mathcal{S}} \bigl(-d(s,\hat{s}_{t-1})\bigr),
\end{equation}
where $d(\cdot,\cdot)$ denotes the spatial distance between two spots and $K$ is a predefined parameter. 
The candidate set at time $t$ is then restricted to
\begin{equation}
\mathcal{C}_t = \mathcal{N}(\hat{s}_{t-1}).
\end{equation}

\subsubsection{Reconstruction-Based Localizer}

\begin{figure}[t]
    \centering
    \includegraphics[width=0.7\linewidth]{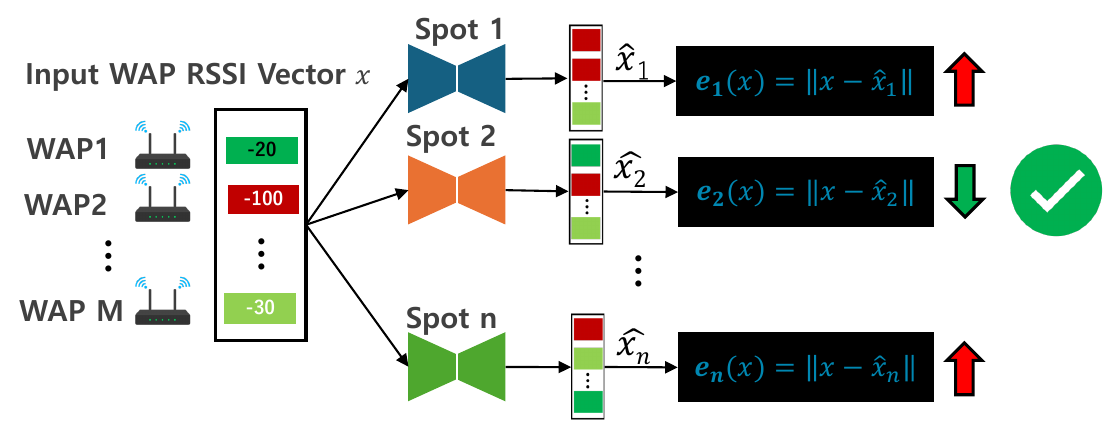}
    \caption{Reconstruction-Error-Based Spot Estimation.}
    \label{fig:inference}
\end{figure}

Given the candidate set selected by HCP or TAP, this module evaluates each candidate local model and estimates the final location using reconstruction error as shown in Figure~\ref{fig:inference}. 
For an input fingerprint $\mathbf{x}$, each candidate model reconstructs the input and produces a reconstructed fingerprint $\hat{\mathbf{x}}_u$. 
The final predicted spot is obtained by selecting the candidate with the minimum reconstruction error:
\begin{equation}
\hat{u} = \arg\min_{u \in \mathcal{C}} e_u(\mathbf{x}),
\end{equation}
where $e_u(\mathbf{x})$ is the masked reconstruction error computed over the observed WAP entries. 
If HCP is used, the final prediction is obtained from the pruned hierarchical candidate set. 
If TAP is used, the final prediction is obtained from the neighborhood of the previously estimated spot.

\section{EXPERIMENTAL SETUP AND EVALUATION}

\subsection{Dataset and Environment}
We evaluate the proposed method using the UJIIndoorLoc~\cite{torres2014ujiindoorloc} dataset, a public wireless local area network (WLAN) fingerprint dataset collected across three buildings of Universitat Jaume I. 
The database covers approximately 108,703~m$^2$ and includes 933 reference points, with measurements captured by more than 20 users and 25 Android devices. 
Each record contains 529 attributes, consisting of 520 RSSI values corresponding to the detected WAPs and 9 metadata fields, namely longitude, latitude, floor, BuildingID, SpaceID\footnote{The space ID is used to define the reference locations inside each building and it is not unique. Therefore, we redefine the mapped space ID of each building to a unique spot ID across all buildings and the entire testbed. }, Relative Position, UserID, PhoneID, and Timestamp. 
The dataset was originally designed for WLAN fingerprint-based indoor localization and provides a realistic multi-building and multi-floor environment for benchmarking.

In the original dataset, RSSI values are represented as integer measurements in the range from -104~dBm to 0~dBm, while unavailable access points are encoded using the value 100.

In this work, we use the 520 RSSI values as the input feature vector to a neural network and train the model to reconstruct the same fingerprint representation.

We do not use the original validation subset of UJIIndoorLoc, since it does not include SpaceID and therefore cannot be directly used for spot-level classification. 
Instead, we split the original training set into training and test subsets with an 8 to 2 ratio. 
The split is performed on a per-spot basis to preserve class coverage and to ensure that each spot retains enough samples for reliable local model training.

Since the dataset does not provide explicit coordinates for spots, the distance between two spots is defined as the Euclidean distance between the centroids of the coordinates of the samples belonging to those spots. 
This definition is used when evaluating distance error and when determining the spatial proximity of candidate spots.

In addition, we build a spot-level dictionary that maps each spot to its building, floor, and representative coordinate. This dictionary is required to compute inter-spot distances and to enable the candidate reduction procedures used in HCP and TAP.

To evaluate history-aware inference, the test samples are further organized into trajectories. 
Specifically, we sort the samples by Timestamp and construct trajectories separately for each UserID and PhoneID pair. 
Continuous samples are then extracted in block form from these trajectories, so that the inference stage can exploit the previously predicted spot when selecting candidate models.

All experiments were conducted on a machine equipped with an NVIDIA L40 GPU (Driver 550.144.03, CUDA 11.8) and an Intel Xeon Gold 6442Y CPU, using Python 3.10.14 and PyTorch 2.3.1. Additionally, to evaluate parallel execution scalability on HPC platforms, multi-node CPU experiments were conducted on the supercomputer Fugaku using multiple nodes equipped with Fujitsu A64FX processors, using Python 3.8.17 and NumPy 1.24.4.

\subsection{Evaluation Metrics and Baseline}
To evaluate the effectiveness of the proposed candidate-pruning methods, we compare them with a brute-force exhaustive baseline following the exhaustive execution strategy adopted in prior modular localization frameworks such as WiDeep~\cite{abbas2019wideep} and CellStory~\cite{saeed2022cellstory}. 
In this baseline, all candidate models are evaluated for each input, and the final prediction is obtained by selecting the candidate with the lowest reconstruction error.
Since the proposed methods differ only in the candidate selection process, this comparison isolates the effect of the proposed candidate-pruning strategy.

For TAP, the previous one-slot location is also obtained using the exhaustive baseline, and the estimated spot from the immediately preceding time step is used as the reference point for candidate reduction. This setting ensures that TAP is evaluated under a realistic online tracking scenario while keeping the comparison with the exhaustive baseline consistent.

To evaluate the proposed framework, we use both localization accuracy metrics and computational efficiency metrics. 
Since the proposed method produces a ranked list of candidate spots, we report top-1, top-3, and top-5 accuracy. 
Top-$k$ accuracy measures the probability that the true Spot ID appears among the $k$ highest-scoring candidates returned by the model. 
This metric reflects whether the correct location is contained in the candidate set selected by the proposed inference procedure.

However, top-$k$ accuracy alone does not distinguish between physically close and distant errors. 
A prediction that is slightly shifted in spot is not equivalent to a prediction that is far away. 
To capture the geometric severity of the error, we also report the localization error in meters. 
In the UJIIndoorLoc dataset, the available spatial coordinates are latitude and longitude, while altitude is not explicitly provided as a continuous geometric quantity. 
Therefore, we additionally evaluate building-level accuracy and floor-level accuracy to measure how often the predicted location is correct at the building and floor levels. 
These metrics allow us to assess both coarse-grained and fine-grained localization performance.

In addition to accuracy, we evaluate the computational cost of inference. 
Since the proposed framework relies on multiple local models, the number of executed models is an important indicator of practical efficiency. 
We therefore report the number of models used to determine the final location estimate.

\subsection{Comparison of Accuracy and Computational Cost}






\begin{figure*}[t]
    \centering

    \begin{subfigure}[b]{0.49\linewidth}
        \centering
        \includegraphics[width=0.9\linewidth]{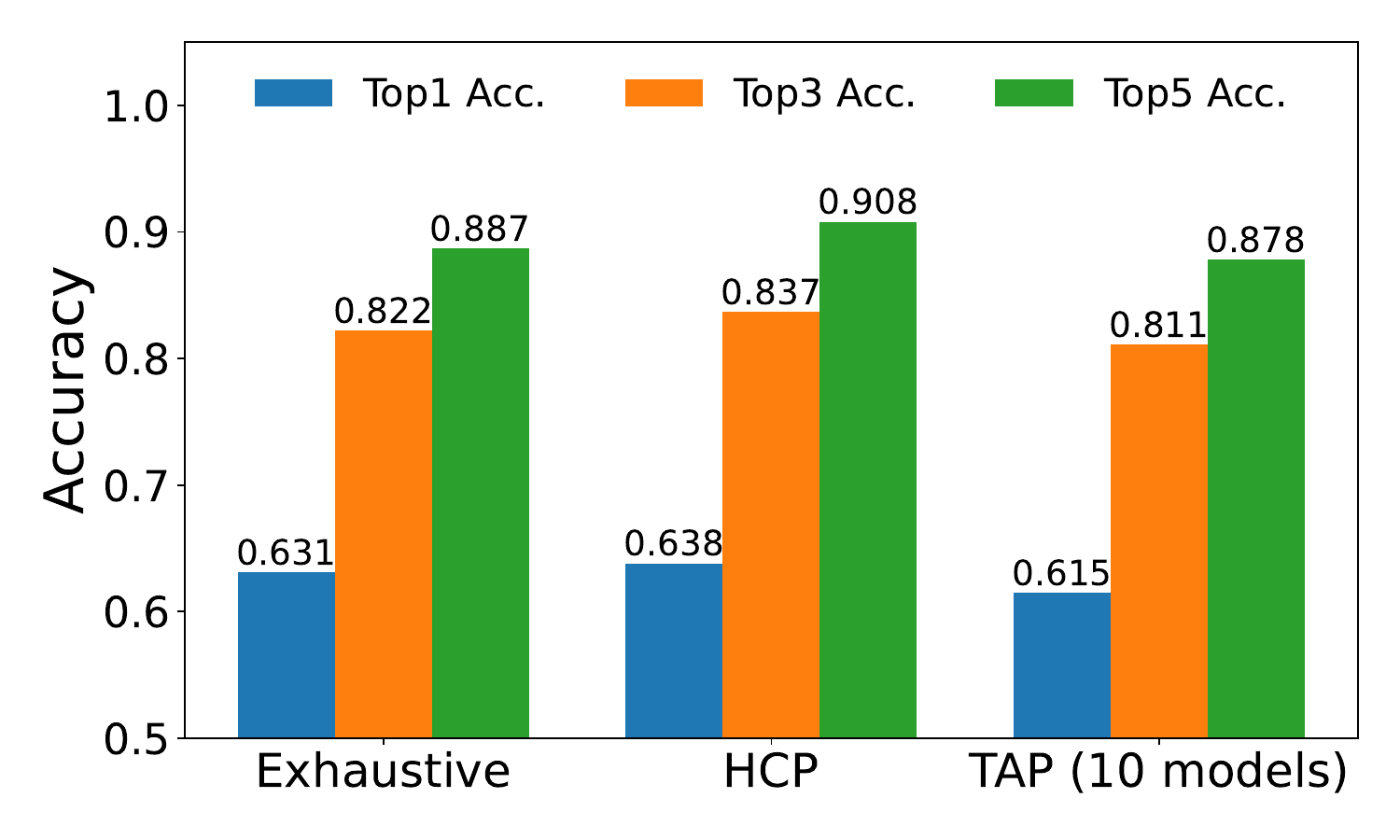}
        \caption{Spot-Level Accuracy}
        \label{fig:spot_acc_comp}
    \end{subfigure}
    \hfill
    \begin{subfigure}[b]{0.49\linewidth}
        \centering
        \includegraphics[width=0.9\linewidth]{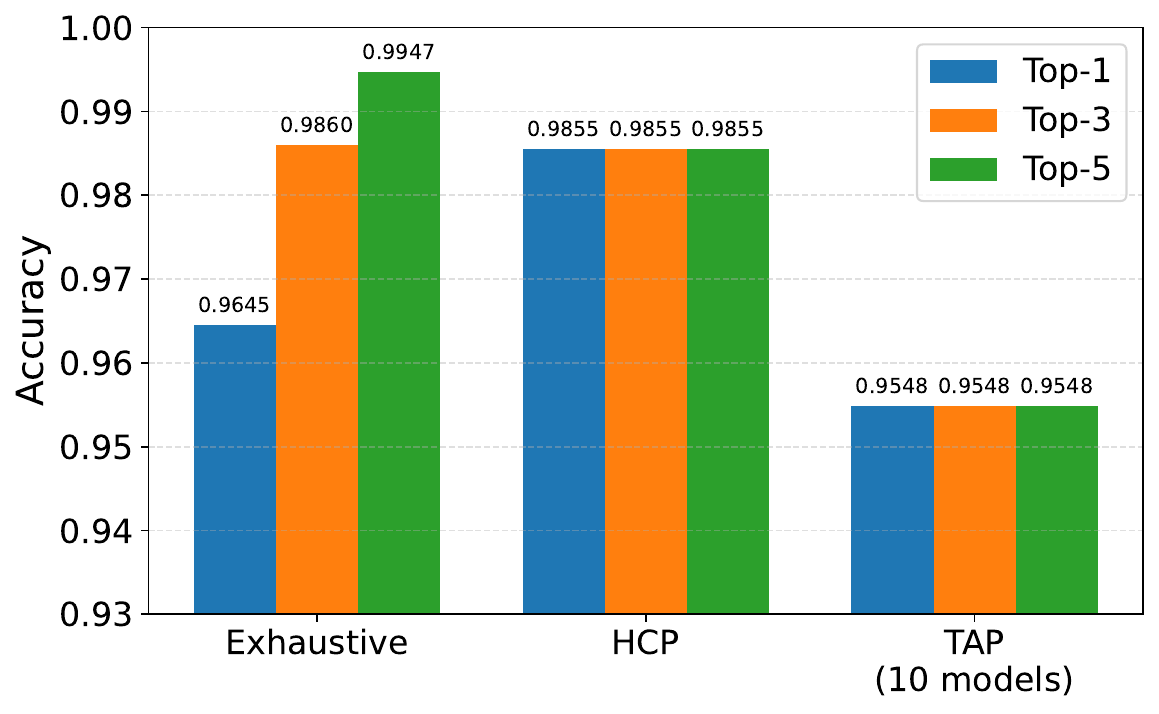}
        \caption{Floor Accuracy}
        \label{fig:floor_acc_comp}
    \end{subfigure}

    \vspace{0.5em}

    \begin{subfigure}[b]{0.49\linewidth}
        \centering
        \includegraphics[width=0.9\linewidth]{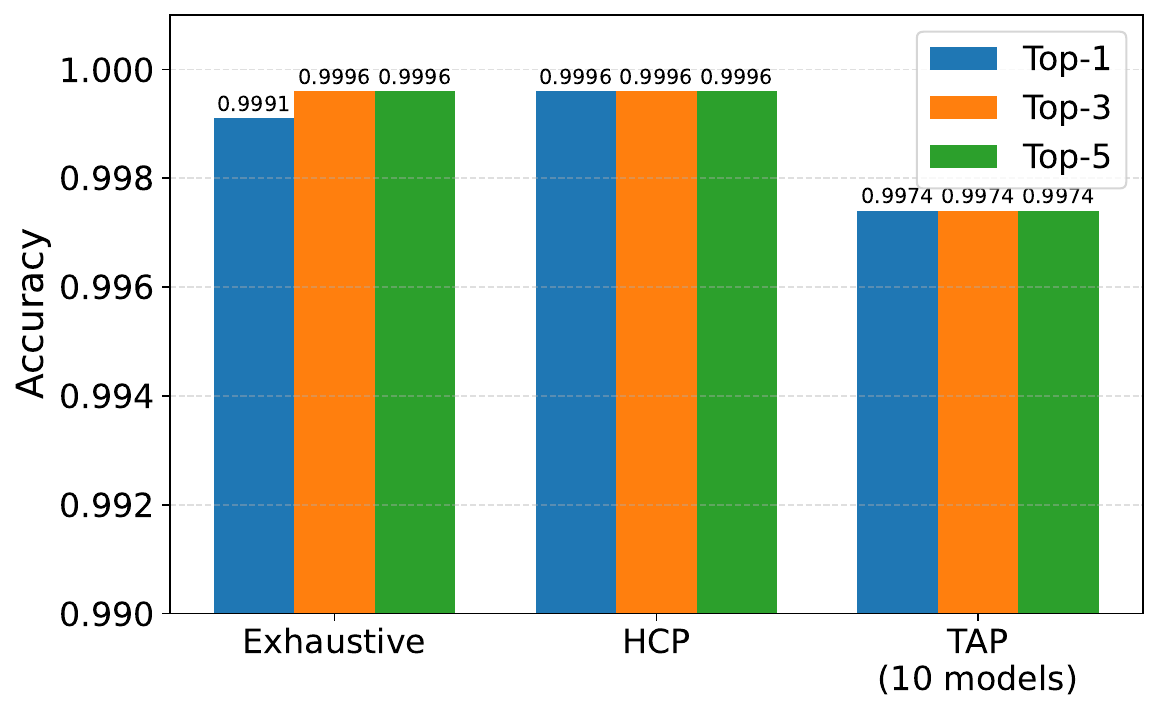}
        \caption{Building Accuracy}
        \label{fig:building_acc_comp}
    \end{subfigure}
    \hfill
    \begin{subfigure}[b]{0.49\linewidth}
        \centering
        \includegraphics[width=0.9\linewidth]{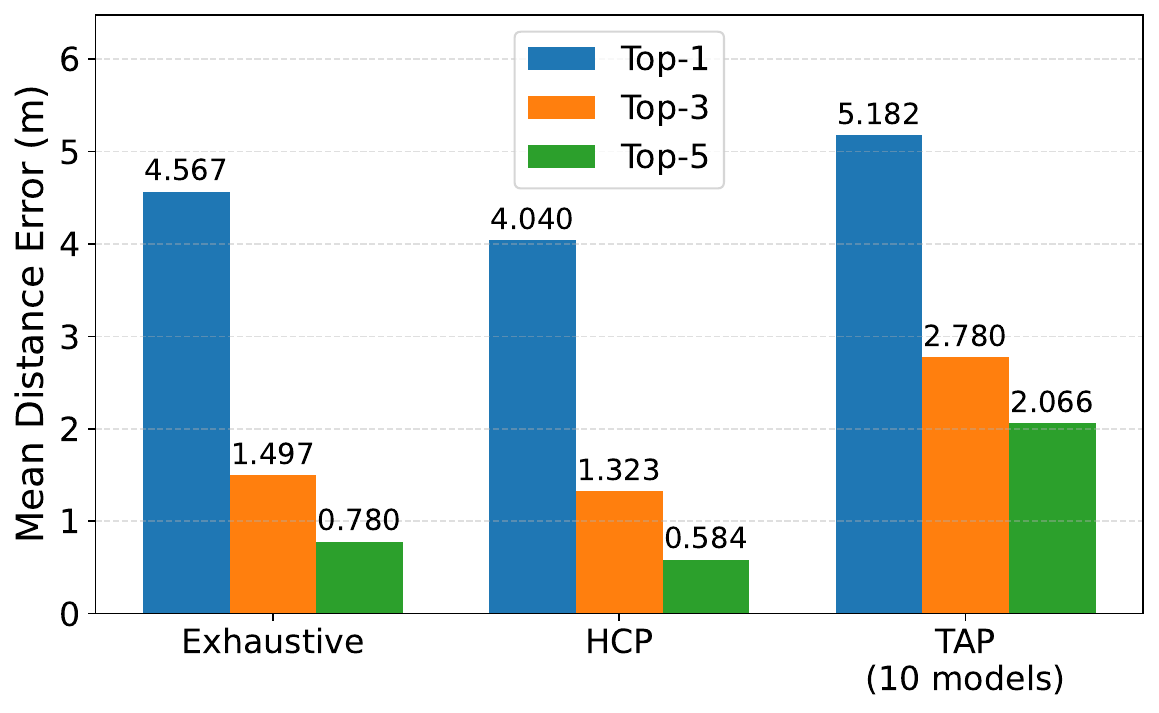}
        \caption{Mean Distance Error}
        \label{fig:comp_dist_error}
    \end{subfigure}

    \vspace{0.5em}

    \begin{subfigure}[b]{0.49\linewidth}
        \centering
        \includegraphics[width=0.9\linewidth]{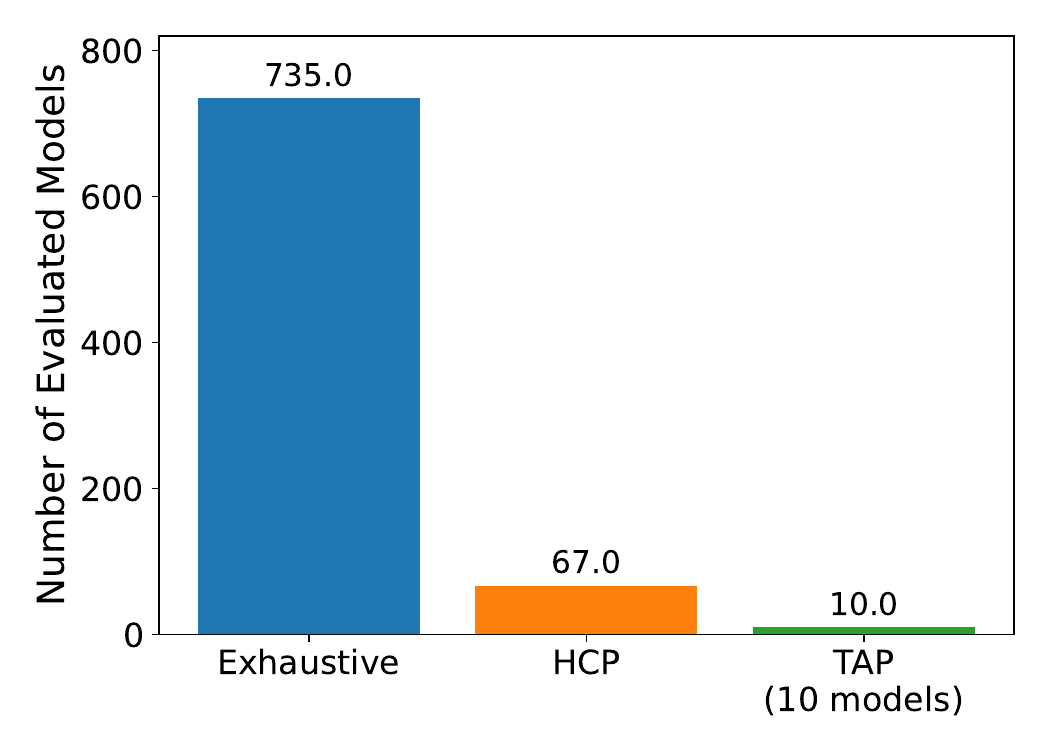}
        \caption{Evaluated Models}
        \label{fig:comp_cost}
    \end{subfigure}

    \caption{Comparison of localization accuracy, distance error, and computational cost.}
    \label{fig:comparison_all}
\end{figure*}

We first compare the conventional exhaustive baseline and the proposed methods in terms of localization accuracy. 
Figure~\ref{fig:spot_acc_comp} reports the top-1, top-3, and top-5 accuracies of the three methods. 
Figure~\ref{fig:comp_dist_error} shows the corresponding distance error, and Figure~\ref{fig:comp_cost} summarizes the required computational cost in terms of the number of executed models.

As shown in Figure~\ref{fig:spot_acc_comp}, the exhaustive baseline achieves a top-1 accuracy of 0.631, while HCP and TAP achieve 0.638 and 0.615, respectively. Overall, the proposed methods maintain localization accuracy at a level comparable to that of the exhaustive baseline, despite requiring substantially fewer model evaluations. In particular, HCP achieves slightly higher top-1, top-3, and top-5 accuracies than the exhaustive baseline, although the differences are relatively small. These results indicate that the hierarchical candidate selection does not degrade localization performance while significantly reducing the computational burden.
This tendency is also reflected in Figure~\ref{fig:comp_dist_error}, where HCP achieves the smallest average distance error of 4.04\,m among the three methods.

In addition to distance errors, we also evaluate the localization performance in the vertical dimension by measuring building-level and floor-level accuracies. 
Figure~\ref{fig:building_acc_comp} presents the building-level accuracy, while Figure~\ref{fig:floor_acc_comp} shows the floor-level accuracy of the three methods. 
As shown in Figure~\ref{fig:building_acc_comp}, all methods achieve nearly perfect building-level accuracy, and the difference among them is negligible. 
This suggests that building-level confusion rarely occurs in the considered dataset. 
A possible explanation is that there are no access points that strongly bridge different buildings, and the physical separation among buildings is sufficiently large to make building discrimination relatively easy.

In contrast, the floor-level results show a more noticeable difference among the methods. 
As shown in Figure~\ref{fig:floor_acc_comp}, HCP achieves the highest floor-level top-1 accuracy of 0.9855. 
This result indicates that explicitly modeling the floor structure helps the system capture floor-specific fingerprint characteristics more effectively than the competing methods. 
In other words, the dedicated floor-level models in HCP provide a better representation of the vertical structure of the environment, which leads to improved floor discrimination.

Figure~\ref{fig:comp_cost} further demonstrates the computational advantage of the proposed methods. The exhaustive baseline requires the execution of 735 models for every localization query. In contrast, TAP achieves comparable localization accuracy while evaluating only 10 models on average per query, corresponding to a 98.6\% reduction in computation. HCP also achieves a substantial reduction, evaluating an average of 67 models per query, consisting of 3.0 building-level models, 4.5 floor-level models, and 59.5 spot-level models. This corresponds to a 90.9\% reduction relative to the exhaustive baseline while maintaining nearly the same localization accuracy. These results confirm that the proposed candidate reduction strategies can significantly improve inference efficiency without sacrificing localization performance. In larger model pools, the advantage of HCP and TAP is expected to become even more pronounced, because the cost of exhaustive evaluation grows with the number of candidate models.

\begin{figure*}[t]
    \centering

    \begin{minipage}{0.49\linewidth}
        \centering
        \includegraphics[width=\linewidth]{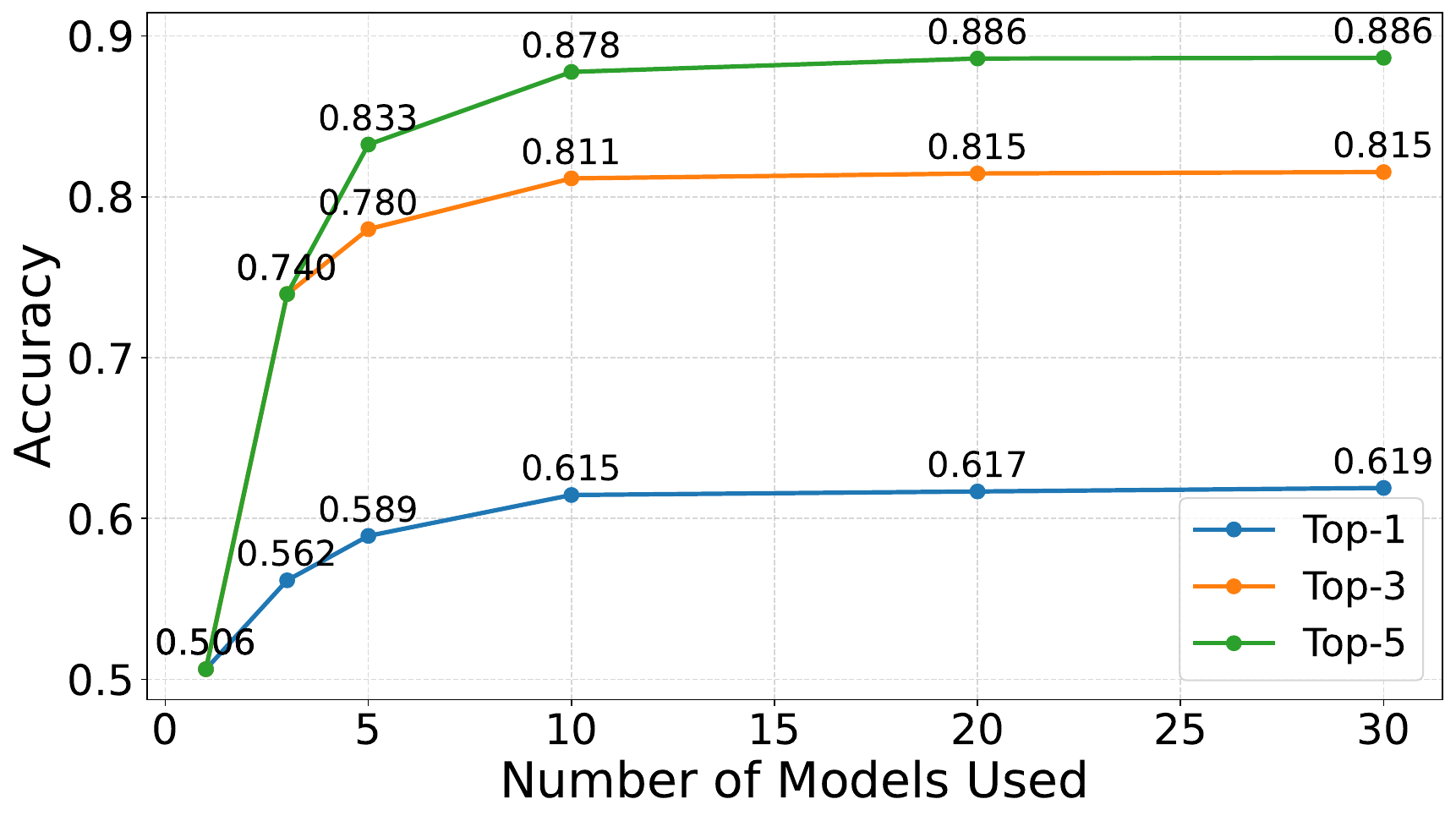}
        \captionof{figure}{Accuracy vs. the Number of Models Selected by TAP.}
        \label{fig:acc_per_num}
    \end{minipage}
    \hfill
    \begin{minipage}{0.49\linewidth}
        \centering
        \includegraphics[width=\linewidth]{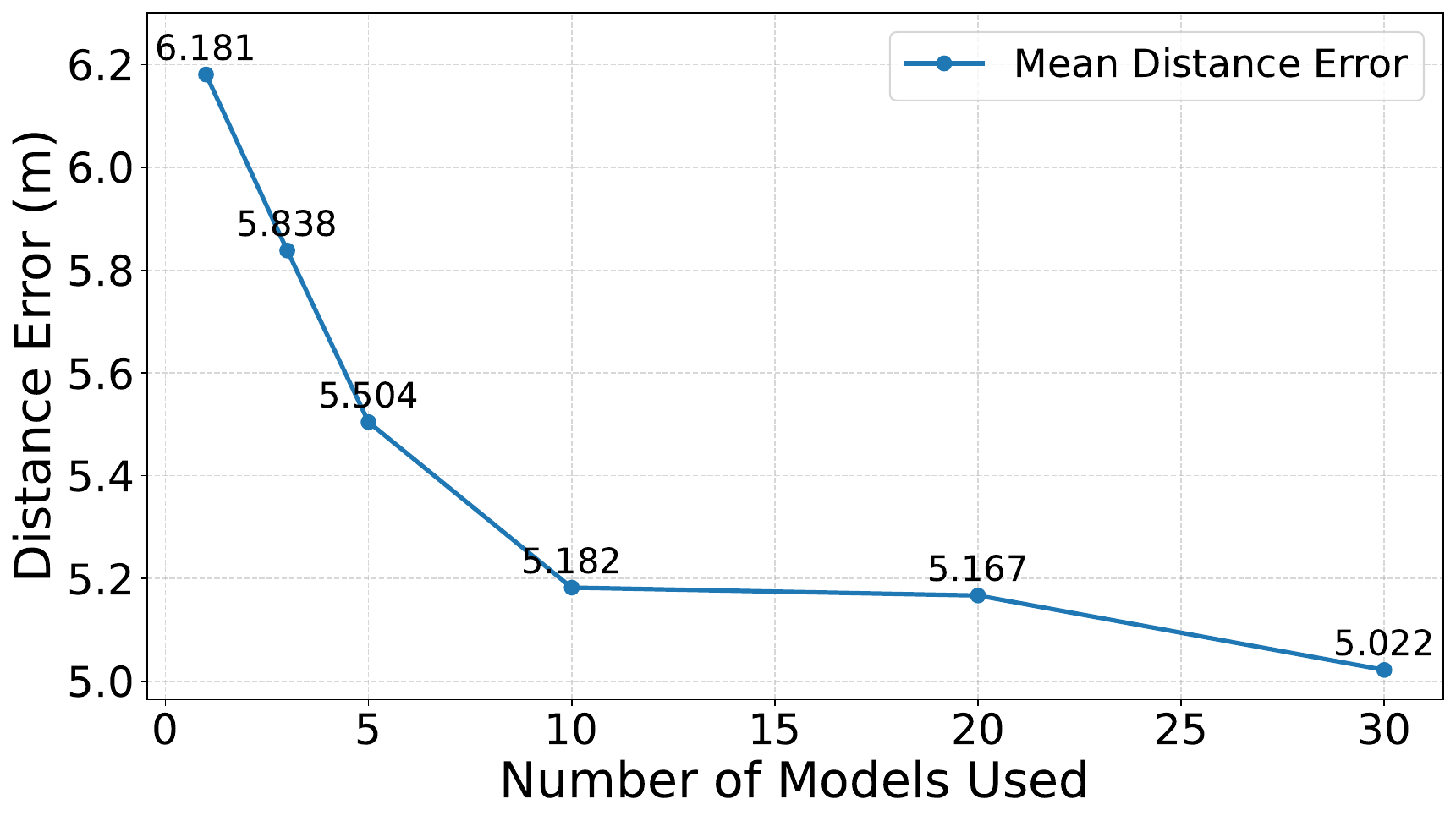}
        \captionof{figure}{Distance Error vs. the Number of Models Selected by TAP.}
        \label{fig:dist_per_num}
    \end{minipage}

    \vspace{0.5cm}

    \begin{minipage}{0.49\linewidth}
        \centering
        \includegraphics[width=\linewidth]{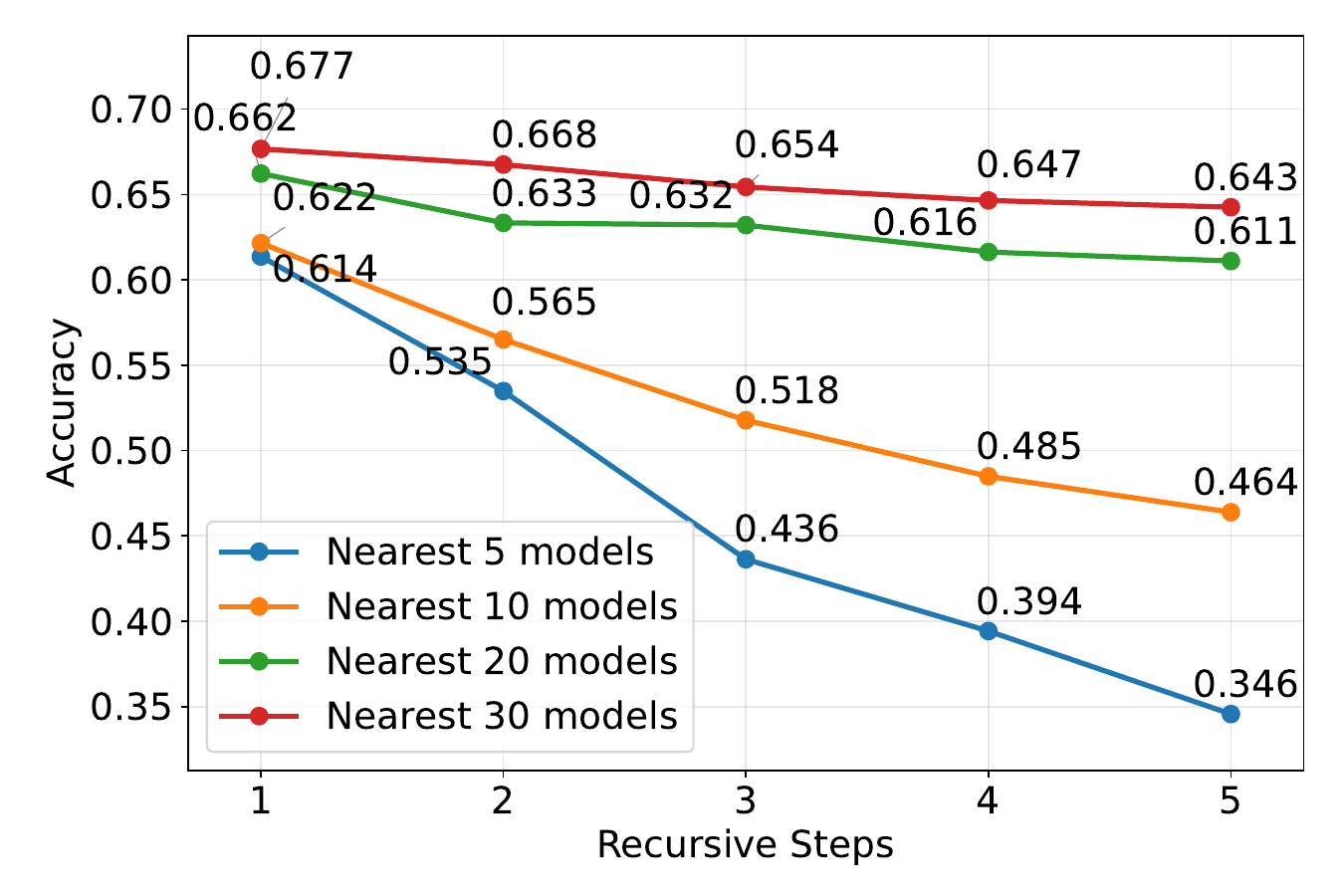}
        \captionof{figure}{Top-1 Accuracy under Recursive TAP.}
        \label{fig:chain_acc}
    \end{minipage}
    \hfill
    \begin{minipage}{0.49\linewidth}
        \centering
        \includegraphics[width=\linewidth]{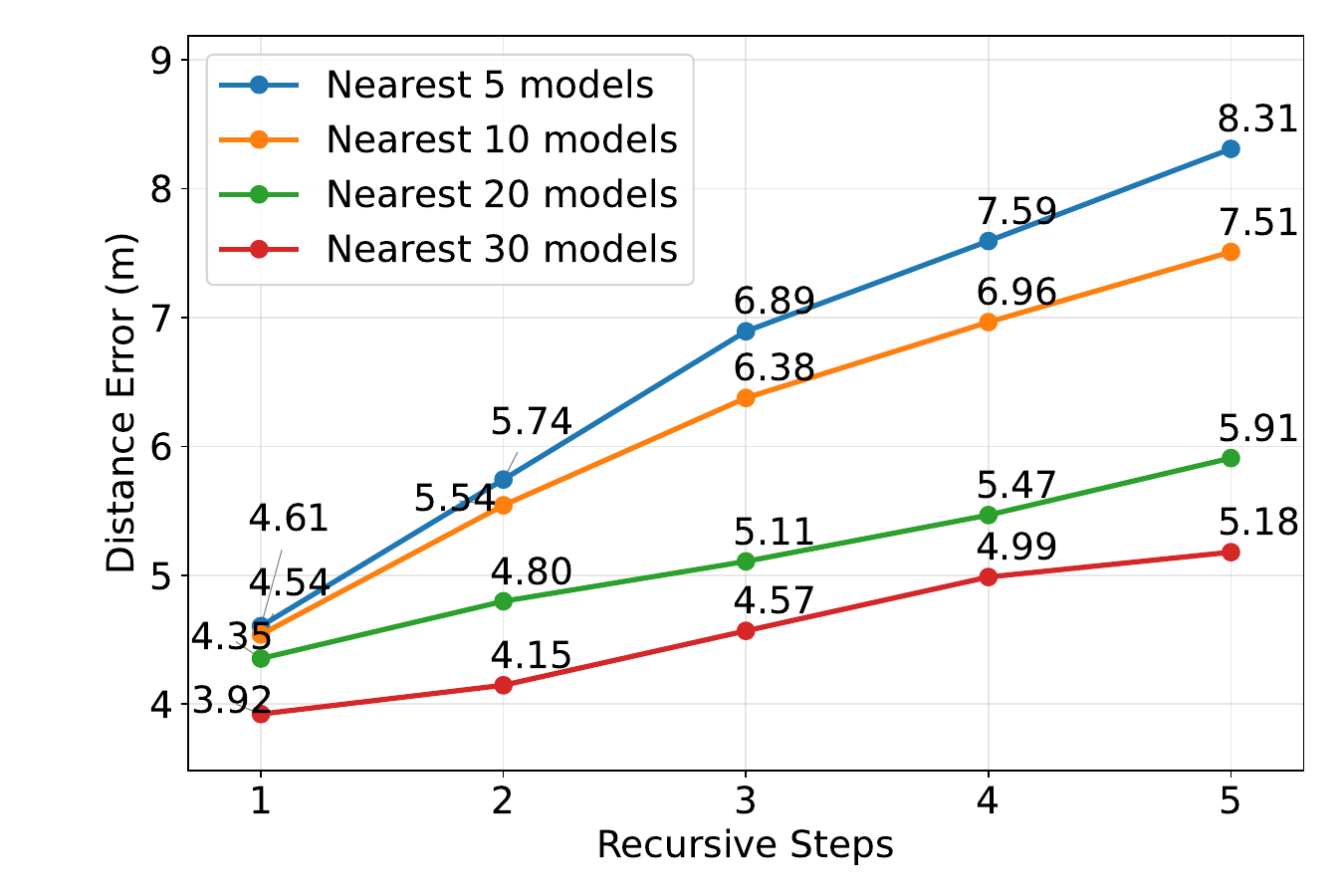}
        \captionof{figure}{Distance Error under Recursive TAP.}
        \label{fig:chain_dist}
    \end{minipage}

    \vspace{0.5cm}

    \begin{minipage}{0.49\linewidth}
        \centering
        \includegraphics[width=\linewidth]{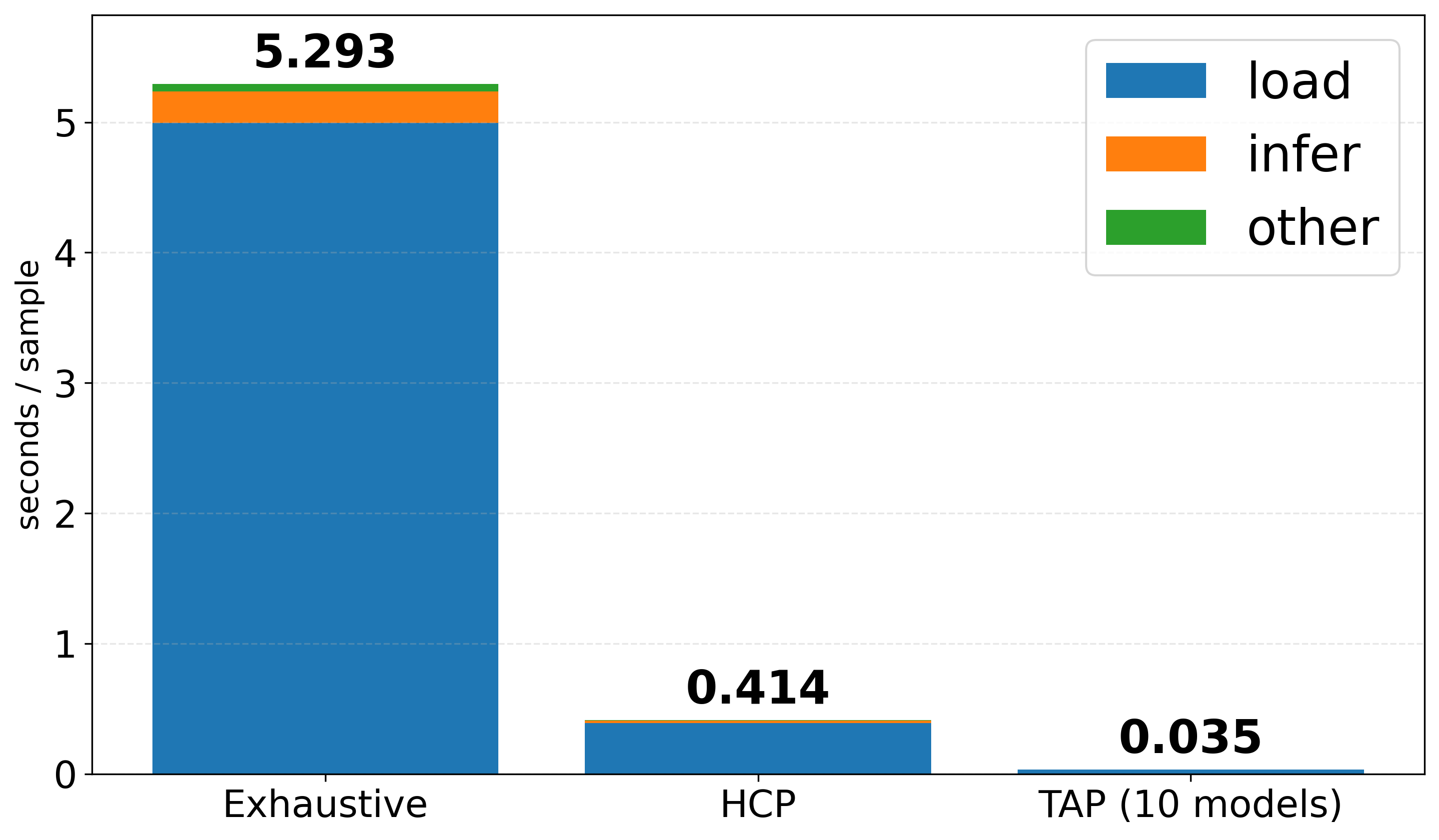}
        \captionof{figure}{Execution Time under a Constrained-Memory Setting.}
        \label{fig:time}
    \end{minipage}
    \hfill
    \begin{minipage}{0.49\linewidth}
        \centering
        \includegraphics[width=\linewidth]{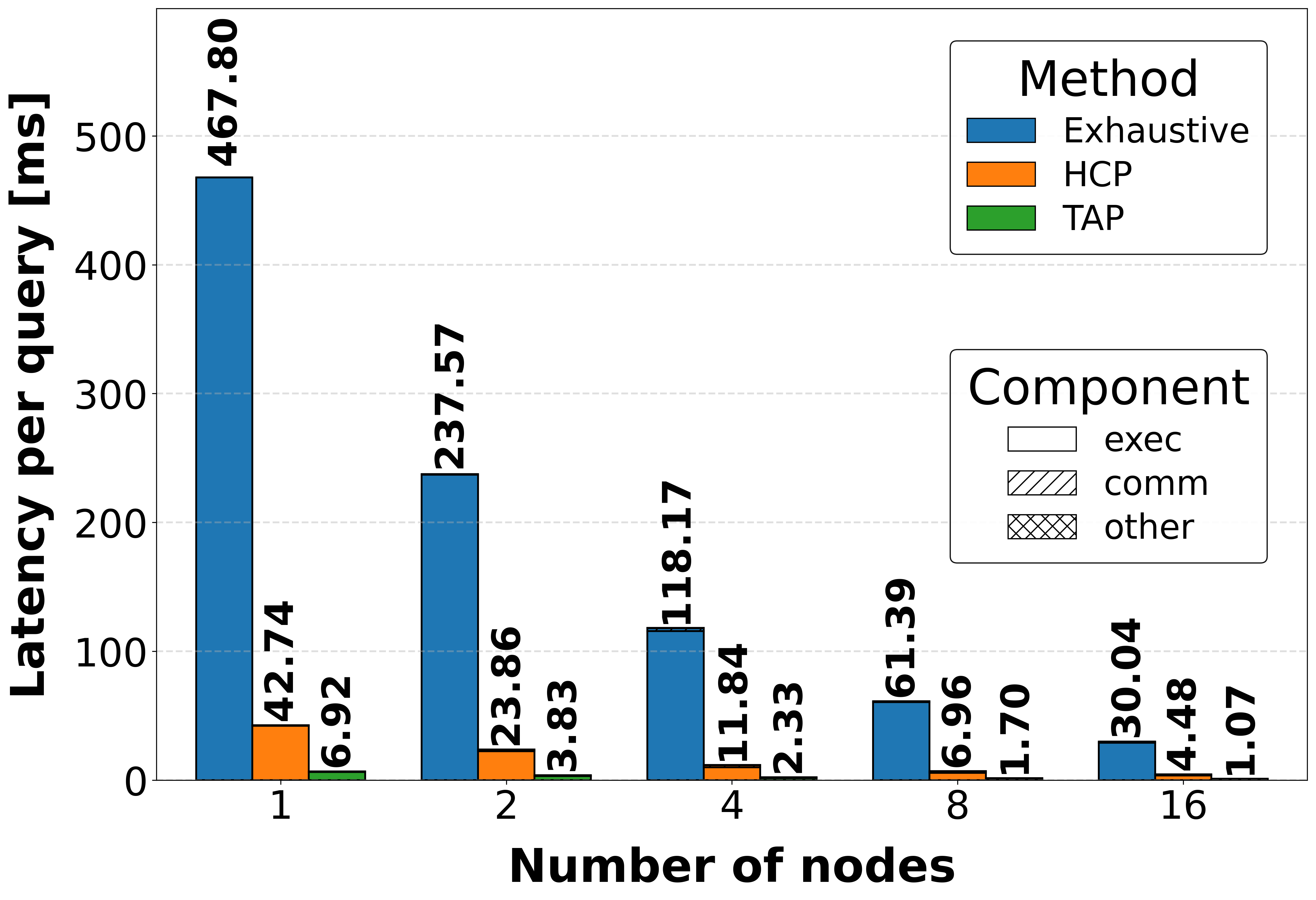}
        \captionof{figure}{Execution Time on Multi-Node CPU.}
        \label{fig:time_hpc}
    \end{minipage}

\end{figure*}

\subsection{Effect of the Number of Neighboring Models in TAP}

We next evaluate the effect of the number of neighboring models used in TAP. 
Figure~\ref{fig:acc_per_num} shows the top-1, top-3, and top-5 accuracies obtained with different numbers of models, and Figure~\ref{fig:dist_per_num} shows the corresponding distance error.

As shown in Figures~\ref{fig:acc_per_num} and~\ref{fig:dist_per_num}, increasing the number of candidate models generally improves localization accuracy. 
This is because the previous estimated spot may contain non-negligible errors, and a larger candidate set can compensate for such uncertainty by increasing the chance that the true location remains within the search region. 
In particular, when only a small number of neighboring models are used, the correct spot may be excluded from the candidate set due to an error in the previous estimate, which directly degrades the final accuracy.

However, the improvement becomes less pronounced after approximately 10 neighboring models in the one-step trajectory-aware setting. 
Beyond this point, the gain in accuracy is marginal, while the computational cost continues to increase. 
Therefore, for the case where TAP relies on the immediately preceding estimate, using about 10 neighboring models provides a reasonable balance between accuracy and efficiency. 
This result indicates that TAP can achieve strong localization performance with a compact candidate set, while avoiding unnecessary model executions in short-term tracking.

\subsection{Robustness Under Recursive Inference}

We also evaluate the robustness of TAP under recursive inference. 
In this experiment, spot estimation and candidate pruning are repeatedly applied, and the effect of increasing recursion depth is examined. 
Specifically, we extract trajectories of length six from the dataset and split them into training and test sets with an 8:2 ratio while preserving the trajectory structure. The models are then trained using the resulting data.

Figures~\ref{fig:chain_acc} and~\ref{fig:chain_dist} show the top-1 accuracy and distance error, respectively, when the number of recursive steps ranges from one to five and the number of neighboring models used in TAP is varied. 
As the number of recursive steps increases, the accuracy gradually decreases and the distance error increases. 
This degradation becomes more pronounced when only a small number of neighboring models, such as five or ten, are used. 
In contrast, when the number of neighboring models is increased to twenty or thirty, the degradation becomes much milder, indicating improved robustness under longer recursive chains.

These results suggest a clear trade-off between computational cost and long-term stability. 
If too few neighboring models are retained, the system becomes more sensitive to accumulated errors in previous predictions. 
In such cases, maintaining accuracy in practical deployment would require more frequent reinitialization through exhaustive search or HCP. 
By contrast, using a larger neighborhood, approximately twenty models in this experiment, provides a better balance between robustness and efficiency.

\subsection{Execution Time Comparison}

Finally, we evaluate the execution-time reduction achieved by the proposed methods across both single-node GPU and multi-node CPU environments.

We first consider a controlled memory-pressure scenario on a single GPU in which the complete model set cannot reside in VRAM simultaneously.
Since increasing the number of models in the actual system is not straightforward, we instead emulate a non-resident execution regime by artificially limiting the available VRAM so that only 100 models can be loaded at a time.
The purpose of this experiment is not to represent a particular GPU configuration, but to investigate the impact of limited model residency on inference latency.
Under this setting, 100 localization queries are processed sequentially using a naive model management scheme that loads the required models on demand and evicts existing models whenever the memory capacity is exceeded, without any explicit VRAM caching strategy.
Figure~\ref{fig:time} shows the average execution time per sample.

As shown in Figure~\ref{fig:time}, when only a subset of models can reside in VRAM, model loading becomes a substantial component of the total execution time.
The exhaustive baseline suffers the largest overhead because it repeatedly loads and executes a large number of models for every localization query.
In contrast, HCP and TAP substantially reduce execution time by decreasing both the number of inference evaluations and model transfers between host memory and VRAM.
If sufficient VRAM is available to keep all models resident, the model-loading overhead becomes negligible and the absolute latency gap is expected to decrease.
Nevertheless, HCP and TAP still require only 67 and 10 model evaluations on average, respectively, compared with 735 for the exhaustive baseline, thereby retaining their computational advantage even without memory pressure.

To further assess parallel scalability in large-scale HPC platforms where all models are fully resident in node memory, we evaluated latency breakdown on the supercomputer Fugaku using 1, 2, 4, 8, and 16 CPU nodes.
Figure~\ref{fig:time_hpc} illustrates the mean latency per query along with its breakdown into model execution (\textit{exec}), inter-node communication (\textit{comm}), and other overheads (\textit{other}).

Across all node configurations, HCP and TAP achieve dramatic latency reductions compared with the exhaustive baseline.
On a single node, mean latency is reduced from $467.80$~ms (exhaustive) to $42.74$~ms (HCP) and $6.92$~ms (TAP).
As the number of compute nodes scales from 1 to 16, latency for the exhaustive baseline decreases almost linearly from $467.80$~ms to $30.04$~ms due to parallelizing the massive model evaluations across nodes.
HCP and TAP scale down to $4.48$~ms and $1.07$~ms at 16 nodes, respectively.

Analyzing the execution breakdown reveals that across all methods and node configurations, model execution (\textit{exec}) consistently dominates the total latency.
Because neural-network evaluations represent the primary computational workload—which becomes even more critical in high-throughput production environments handling concurrent user queries—substantially compressing \textit{exec} is key to scalable execution.
These results demonstrate that reducing the required number of model evaluations via HCP and TAP is universally effective regardless of the computing architecture (GPU vs. CPU) or memory conditions, delivering high scalability and throughput for parallel HPC inference workloads.

\section{Conclusion}
In this paper, we proposed a modular indoor localization framework that treats localization as a fine-grained spot estimation problem rather than a single monolithic classification task. 
The proposed method represents the environment as a collection of local models defined at the building, floor, and spot levels, and it combines two complementary candidate reduction strategies, namely hierarchical candidate pruning and trajectory-aware pruning. 
HCP exploits the spatial hierarchy of the environment by first selecting the most likely building, then the most likely floor, and finally the most likely spot. 
TAP further reduces the search spot by leveraging temporal continuity and restricting the candidate spots to the neighborhood of the previously estimated location. 
These design choices make the framework more suitable for large-scale and long-term deployment than conventional exhaustive fingerprint-based approaches.

The experimental results on the UJIIndoorLoc dataset demonstrate that the proposed framework can substantially reduce inference cost while maintaining localization performance. Compared with the exhaustive baseline, HCP and TAP achieved comparable spot-level localization accuracy while significantly reducing the number of executed models. In particular, the exhaustive baseline required 735 model evaluations per query, whereas HCP and TAP reduced this number to 67 and 10, respectively, corresponding to reductions of 90.9\% and 98.6\%. Despite this large reduction in computational cost, the degradation in top-1 accuracy remained small. Furthermore, HCP achieved the lowest mean distance error and the highest floor-level accuracy, indicating that explicitly modeling the building--floor--spot hierarchy effectively captures the spatial structure of indoor environments. 

As future work, we plan to investigate cache management strategies under multi-user and multi-query settings, as well as resource allocation policies for high-performance computing-based execution environments. 

\bibliographystyle{IEEEtran}
\bibliography{reference}

\end{document}